\documentclass[aps,preprint]{revtex4}%
\usepackage{amsmath}
\usepackage{graphicx}%
\usepackage{amsfonts}%
\usepackage{amssymb}
\begin{document}

\title{Numerical evidences of spin-1/2 chain approaching spin-1 chain}
\author{Hsiang-Hsuan Hung$^{1}$ and Chang-De Gong$^{2,1,3}$}
\affiliation{$^{1}$National Key Laboratory of Solid State Microstructure and Department of
Physics, Nanjing University, Nanjing 210093, China}
\affiliation{$^{2}$Chinese Center of Advanced Science and Technology (World Laboratory)
P.O. Box 8730, Beijing 100080, China}
\affiliation{$^{3}$Department of Physics, The Chinese University of Hong Kong, Hong Kong, China}

\begin{abstract}
\textit{In this article, we study the one dimensional Heisenberg spin-1/2
alternating bond chain in which the nearest neighbor exchange couplings are
ferromagnetic (FM) and antiferromagnetic (AF) alternatively. By using exact
diagonalization and density matrix renormalization groups (DMRG) method, we
discuss how the system approaches to the AF uniform spin-1 chain under certain
condition. When the ratio of AF to FM coupling strength }$\alpha$
$(\alpha=J_{AF}/J_{F})$ \textit{is very small, the physical quantities of the
alternating bond chain such as the spin-spin correlation, the string
correlation function and the spin density coincide with that of the AF uniform
spin-1 chain. The edge state problem is discussed in the present model with
small }$\alpha$\textit{\ limit. In addition, the Haldane gap of the AF uniform
spin-1 chain is 4-times of the gap of the system considered.}

\end{abstract}
\maketitle

\address{} 

Haldane argued that the elementary excitation spectrum in the one dimensional
(1-D) antiferromagnetic Heisenberg (AFH) spin system with integer spin is
gapful (massive) whereas for the system with half-integer spin is gapless
(massless).\cite{F.D.M.Haldane} Furthermore, the spin-spin correlation in the
ground state for 1-D AFH spin-1 systems should have the exponential decay
behavior due to the gapful excitation energy. By contrary, in 1-D AFH
spin-$1/2$ systems, the spin-spin correlation function exhibits power law
decay behavior due to the gapless excitation energy. There are many numerical
studies to investigate this conjecture.\cite{M.P.Nightingale et al.}%
\cite{S.R.White and Huse}\cite{T.Kennedy}\cite{Golinelli et al.}\cite{Ulrich
Schollwock} On the other hand, the spin gap has been observed in the compound
NENP[Ni(C$_{2}$H$_{8}$N$_{2}$)$_{2}$NO$_{2}$ClO$_{4}$] through
the\ susceptibility measurement by inelastic neutron scattering
experiment.\cite{J.P.Renard et al.}

Among different studies of the Haldane problem in AFH spin-1 chain, the
interested one is given by AKLT model.\cite{AKLT} \ This model gave a simple
picture for the ground state of 1-D AFH spin-1 system. Each S=1 spin was
viewed as a triplet state of two spins with $S=1/2$. According to this
picture, only two end spins of the open AKLT chain are left and form two
''free''\ $S=1/2$ spins which are called the edge state and all the other
$S=1/2$ spins form RVB type spin singlet state between nearest neighbor spins.
In the thermodynamic limit, the spin-1 chain is fourfold degeneracy due to the
two $S=1/2$ spins. This picture was observed experimentally\cite{S.H.Glarum et
al}\cite{M.Hagiwara} and discussed theoretically.\cite{T.Kennedy}\cite{Shaojin
Qin}\ For an open spin-1 chain the Haldane gap is the difference between the
lowest eigen-energy in subspace $S_{tot}^{z}=0$ (or $S_{tot}^{z}=1$) and
$S_{tot}^{z}=2.$ Although spin-1 chain is a disorder system, it still has a
hidden order, the corresponding order parameter proposed by M.den Nijs and K.
Rommelse\cite{Nijs et al.} is represented by the string correlation defined in
the later discussion. Since after, the numerical studies have been performed
for 1-D spin-1 Heisenberg model and have proven that the systems really posses
the hidden order.\cite{S.M.Girvin et al.}\cite{Alcaeaz et al.}

In this article the Haldane problem is studied from another reversed approach.
We start from a 1-D AFH spin chain with $S=1/2$, the exchange couplings
between nearest neighbor spins take $J_{1}$ and $J_{2}$ alternatively as shown
in Fig.1. The Hamiltonian is%
\begin{equation}
H=J_{1}\underset{<i\in odd>}{\sum}\overrightarrow{S}_{i}\cdot\overrightarrow
{S}_{i+1}+J_{2}\underset{<j\in even>}{\sum}\overrightarrow{S}_{j}%
\cdot\overrightarrow{S}_{j+1}%
\end{equation}
This model which is called alternating bond chain has been studied in
detail.\cite{J.C.Bonner et al.} In general, the excitation spectrum of this
model is gapful except the case of $J_{1}=J_{2}$, i.e. it reduced to the
uniform spin-$1/2$ chain. We may ask the questions that, what is the relation
between the Haldane gap and the gap of model (1), and how can we understand
the edge state of AKLT model in a reasonable way. For definiteness, we are
interested in the special case, where $J_{1}<0$ (ferromagnetic coupling FM)
and $J_{2}>0$ (antiferromagnetic coupling AF). We define a parameter
$\alpha=|J_{2}/J_{1}|$ and explore how the magnetic behavior of the system
evolutes with the parameter $\alpha$. We use two numerical techniques
including the exact diagonalization method up to the system N=32 (N is the
number of sites.) with periodic boundary condition and DMRG
method\cite{S.R.White} to even more large system (N=360) with open boundary condition.

At first, in the case of large $\alpha$ limit, the two nearest neighbor spins
connected with AF bond will form an AF singlet dimer and the 1-D chain reduces
to $N/2$ S=0 dimers connected with FM bond. As well known that the first
excited state is the state in which one of the N S=0 dimers is excited to
become a triplet, i.e. the excitation energy is $\Delta=J_{2}$. In opposite
case of small $\alpha$ limit, the two nearest neighbor spins connected with a
FM bond will form a FM triplet dimer and the AF coupling exists between
triplet dimers. Thus we can image that the system will behave like an AFH
uniform spin-1 chain. In the following discussion we will argue this
imagination through some physical properties calculation for Hamiltonian (1)
in the small $\alpha$ cases and compare them with corresponding properties of
the 1-D AFH spin-1 chain.

We find that all the properties for the two systems coincide with each other
in the small $\alpha$ limit. In Fig.2, the exact diagonalization result of the
model (1) for the spin-spin correlation on 32 sites is shown by solid circles.
The plateaus in the spin-spin correlation occur at the pairs of site. Within
each pair the spins couple each other with the FM interaction. This means that
in small $\alpha$ limit each pair of spins ($S=1/2$) corresponds to a FM
triplet dimer as mentioned above. The spin-spin correlation of the AF uniform
spin-1 chain on 16 sites is also shown in the same figure for comparison. (The
data of spin-1 chain should be reduced by a factor 4 in comparison with the
data of model (1).\cite{Golinelli et al.}) It can be seen that when
$\alpha=0.01$ the locus of plateaus of model (1) approaches to that of the AF
uniform spin-1 chain.

Beside the spin-spin correlation, the string correlation of the alternating
bond chain also shows Haldane-like behavior. We define the string correlation
of model (1) as follows%

\begin{equation}
\ \vartheta_{alt}^{z}(j)=\left\langle (S_{1}^{z}+S_{2}^{z})\exp[i\pi
\underset{k=1}{\overset{j-1}{\sum}}(S_{2k-1}^{z}+S_{2k}^{z})](S_{2j-1}%
^{z}+S_{2j}^{z})\right\rangle
\end{equation}
where $\langle..\rangle$ means taking the average over the ground state of the
model considered. This is a simple generalization of the definition of the
string correlation of integer spin S chain\cite{S.M.Girvin et al.}%
$\ \ \ \ \ \ \ \ \ \ \ \ \ \ \ \ \ \ \ \ \ \ \ \ \ $%
\begin{equation}
\ \vartheta_{\pi/S}^{z}(j)=\left\langle S_{1}^{z}\exp[i\frac{\pi}{s}%
\overset{j-1}{\underset{k=1}{\sum}}S_{k}^{z}]S_{j}^{z}\right\rangle
\end{equation}
In considering the two spins with FM coupling in a $S=1/2$ alternating bond
chain will form a triplet dimer in small $\alpha$\ limit, which is
corresponding to a spin-1 chain. The numerical results of the string
correlation for two models are shown in Fig.3. We can see that upon decreasing
$\alpha$, the string correlation of the alternating bond chain is closer and
closer to that of the AF uniform spin-1 chain.

In addition to the spin-spin and the string correlations, the average of
z-component of spin at each site $\langle S_{i}^{z}\rangle$ of the alternating
bond chain will also show the ''edge state''\ with average spin at the ''end
sites''\ proposed by AKLT model. In the following discussion we will
understand what do the ''edge state''\ and ''end sites''\ mean in our
approach: From $S=1/2$ model (1) to an uniform spin-1 chain. We now calculate
the $\langle S_{i}^{z}\rangle$ of model (1) in subspace $S_{tot}^{z}=1$ by
using DMRG method and consider open boundary condition for both systems. In
Fig.4 (a) the results are shown for the system with N=60 lattice sites and
$\alpha=0.01$. It is clear that the two spins with the FM coupling are pairing
to triplet dimers, while the coupling between nearest neighbor dimers is AF.
The average value $\langle S_{i}^{z}\rangle$ for the two spins in each dimer
are nearly the same, e.g. each spin of the edge dimer gives $\langle
S_{1,2}^{z}\rangle\simeq0.27$. Upon entering into the interior of the chain,
the $\langle S_{i}^{z}\rangle$ decreases. At the two centre dimers, the
$\langle S_{i}^{z}\rangle$ are zero. Increasing the size of the chain to
N=120, we can see from Fig.4(b) that: i) the average value of $\langle
S_{i}^{z}\rangle$ in each dimer is nearly equal to the corresponding one in
the case of N=60, and takes nearly the same value irrespective to
$\alpha=0.01$ or $\alpha=0.001$. The loci of dimers are well-coincident with
the $\langle S_{i}^{z}\rangle/2$ of the uniform spin-1 system, e.g. for
$\alpha=0.001$, the $\langle S_{1}^{z}+S_{2}^{z}\rangle\sim0.531$ for the edge
dimer of model (1), it's considerably close to the value (0.532) of the end
spin in the AFH spin-1 chain calculated by S.R.White.\cite{S.R.White} \ ii)
more dimers of spins located in the middle part of the chain with $\langle
S_{i}^{z}\rangle\simeq0$. From these results, we conclude that the ''edge
state''\ in the AFH spin-1 system is actually the state of edge dimer in model (1).

In addition, from Fig.4(c), we can see that as the N increases further, such
as N=240 and 360, the two side regions where the $\langle S_{i}^{z}\rangle
\neq0$ change little because they come from the ''surface''\ effect. So, we
may image that, in the limit of $N\rightarrow\infty$, $N_{0}/N\rightarrow1$.
($N_{0}$ is the number of sites in the middle part.) The effect is due to the
spin disturbance which is happened at the surface (surface mode) and the
penetration depth $\xi$ of the disturbance is independent of the system size,
but it will depend on $\alpha$. The decay behavior of $\langle S_{i}%
^{z}\rangle$ away from the end point of the chain in model (1) as shown in
Fig.4(c) can be fitted by the exponential form $|\langle S_{i}^{z}%
\rangle|=|\langle S_{0}^{z}\rangle|e^{-\frac{i}{\xi}}$. In the small $\alpha$
region, $\xi$ decreases upon increasing $\alpha$ and the relation between
$\xi$ and $\alpha$ is approximately linear. Through numerical calculations, we
could conclude that $\xi\simeq12$ as $N\rightarrow\infty$ and $\alpha
\rightarrow0$. The penetration depth $\xi$ is approximately the double of the
decay length of the AFH spin-1 chain\cite{S.R.White and Huse}\cite{Golinelli
et al.} and this is because two spins connected with a FM coupling of model
(1) correspond to a S=1 spin of the AFH spin-1 chain. These conclusions
support AKLT model but it should be noted that: i) the edge state in which the
average value of end spin equal to nearly $S/2$ is valid for both the AFH
uniform spin-1 system and $S=1/2$ system of model (1) as state above, i.e. it
is not necessary explained by AKLT model in general. ii) The side region,
originating from ''surface'' effect, is finite and independent of the system size.

At last, we turn to study the excitation energy $\Delta_{1/2}$ of model (1) in
small $\alpha$ limit, $\Delta_{1/2}$ is defined as $\Delta_{1/2}=E_{1}-E_{0}$,
where $E_{0}$ and $E_{1}$ are the energies of ground state and the lowest
excite state of model (1) with periodic boundary condition respectively. For
open boundary condition, one must define the gap $\Delta_{1/2}$ $(\Delta_{1})$
of the alternating bond chain (the uniform spin-1 chain)\ as the difference
between the lowest-lying states with $S_{T}=2$ and $S_{T}=0$ (or, $S_{T}=2$
and $S_{T}=1$).\cite{T.Kennedy}\cite{S.R.White} As a matter of fact, the two
gaps of model (1): $\Delta_{1/2}=E_{2}-E_{0}$ and $\Delta_{1/2}=E_{2}-E_{1}$
are equal. The $\Delta$ for both the $S=1/2$ model (1) with $J_{1}=1$,
$J_{2}=\alpha=0.002$ and the AFH spin-1 system with the same $J_{2}$ (i.e. The
Hamiltonian of the spin-1 chain is $H=J\underset{i}{%
{\textstyle\sum}
}\overrightarrow{S}_{i}\cdot\overrightarrow{S}_{i+1}$, $\ $here $J=J_{2}$) are
shown in Fig.5. In the thermodynamic limit ($N\rightarrow\infty$), the
excitations of both systems through above definition are gapful. (For the
spin-1 chain, $\Delta_{1}=\Delta_{Haldane}J,$ $\Delta_{Haldane}\sim
0.41$.\cite{S.R.White and Huse}\cite{Golinelli et al.}) This is the so called
Haldane gap and, however, the gap of the uniform spin-1 system is nearly
4-times of the gap of model (1). If we recalculate $\Delta_{1}$ for the AF
uniform spin-1 system with $J=J_{2}=\alpha/4$ and its locus is denoted by
stars, then we find that they nearly complete coincide with the corresponding
locus of model (1). The appearance of factor 4 is due to the following reason.
The ground state energy of model (1) is
\begin{equation}
E_{0}=\langle0|H|0\rangle=J_{1}\underset{<i\in odd>}{\sum}\langle
\overrightarrow{S}_{i}\cdot\overrightarrow{S}_{i+1}\rangle_{0}+J_{2}%
\underset{<j\in even>}{\sum}\langle\overrightarrow{S}_{j}\cdot\overrightarrow
{S}_{j+1}\rangle_{0}%
\end{equation}
where $\left|  0\right\rangle $ is the ground state.

From Fig.6(a) we can see that the mean value of local bond strength $\langle
S_{i}\cdot S_{i+1}\rangle_{S=1/2}$ of all pairs of spins coupled with a FM
interaction equal to 0.25 and contributes an energy $0.25J_{1}$. In same, the
mean value of all pairs of spins coupled with a AF interaction is equal to
$-0.35$ and contributes an energy of $0.35J_{2}$. In small $\alpha$ limit, all
the pairs with FM coupling form the triplet dimers, and between the dimers,
there exist weak AF interactions. Now let us consider the lowest excite state.
It is obviously that if we change the spin state of a triplet dimer as a
whole, the energy cost is about $4\times0.35J_{2}$. This is much lower than
the excitation energy in which only single one spin state is changed. This is
about $2(0.25J_{1}+0.35J_{2})$. So the gap of model (1) should be
approximately written as $\Delta_{1/2}\simeq4\times0.35J_{2}$. In similar
consideration, the excitation energy of the AF uniform spin-1 system is
$\Delta_{1}\simeq4J\langle S_{i}\cdot S_{i+1}\rangle_{S=1}=4\times1.40J_{2}$
(see Fig.6(b)\cite{S.R.White}). Therefore we have got the relation $\Delta
_{1}=4\Delta_{1/2}$ or equivalently the $J_{2}$ in the spin-1 system
effectively equals to 4 times of $J_{2}$ in model (1). In addition, like the
AFH spin-1 systems, the excitation of model (1) between the lowest-lying
$S_{T}=1$ and $S_{T}=0$ states is gapless due to the four-fold degenerate
ground state generated by two free spins of ends of systems.\cite{S.R.White
and Huse}\cite{T.Kennedy}\cite{S.R.White}

In summary, we presented numerical results of model (1) and compare with the
AFH uniform spin-1 chain. From the spin-spin correlation and the string
correlation, the system with small $\alpha$\ has Haldane-like behavior. The
plateaus of the spin-spin correlation and the spin density of model (1) with
small $\alpha$\ display that spins coupled with a FM interaction form a FM
triplet dimer. The edge state of model (1) is also observed and $\langle
S_{1}^{z}+S_{2}^{z}\rangle$ of model (1) is nearly equal to $\langle S_{1}%
^{z}\rangle$ of the AFH spin-1 chain. Furthermore, the penetration depth $\xi$
of model (1) is the double of the decay length of the AFH spin-1 chain and
then the lattice constant of AFH spin-1 chain is equivalent to the double of
that of model (1). In addition, we find that the gap of model (1) with small
$\alpha$ is not only finite but it's also a quarter of the Haldane gap of the
AFH uniform spin-1 chain. This is because the energy cost for changing the
state of the triplet dimer as a whole is less than that for changing only one
spin of the triplet dimer and the local bond strength of the AFH uniform
spin-1 chain is 4-times of that of model (1).

\begin{acknowledgments}
The authors are grateful for the support from the Ministry of Science and
Technology of China under Grant No. NKBRSF-G19990646
\end{acknowledgments}

\bigskip

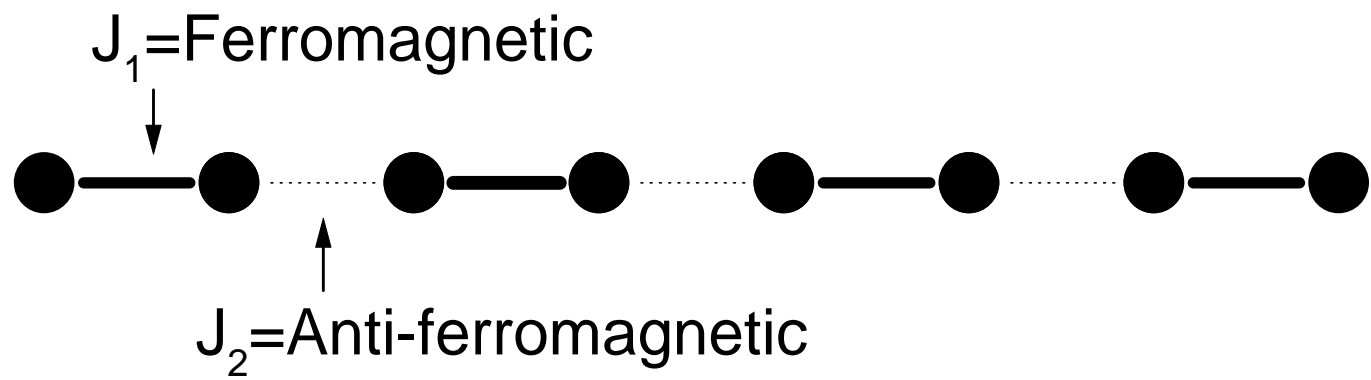
Figure 1: The structure of an alternating bond chain.

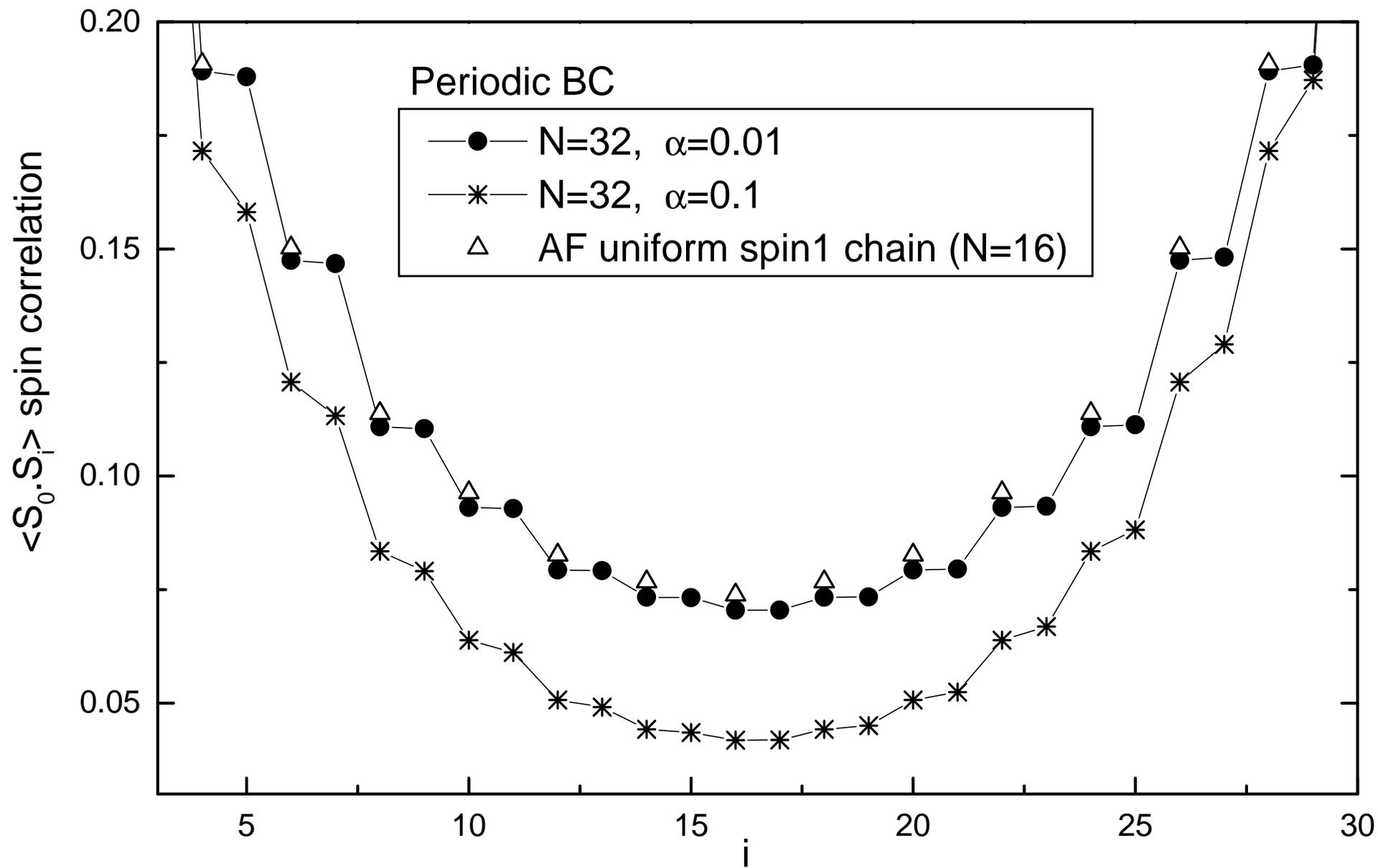
Figure 2: The spin-spin correlation of the ground state of the alternating
bond chain (N=32) with periodic boundary condition.\ It shows many plateaus.
In each plateau, the two spins have the same spin correlations.

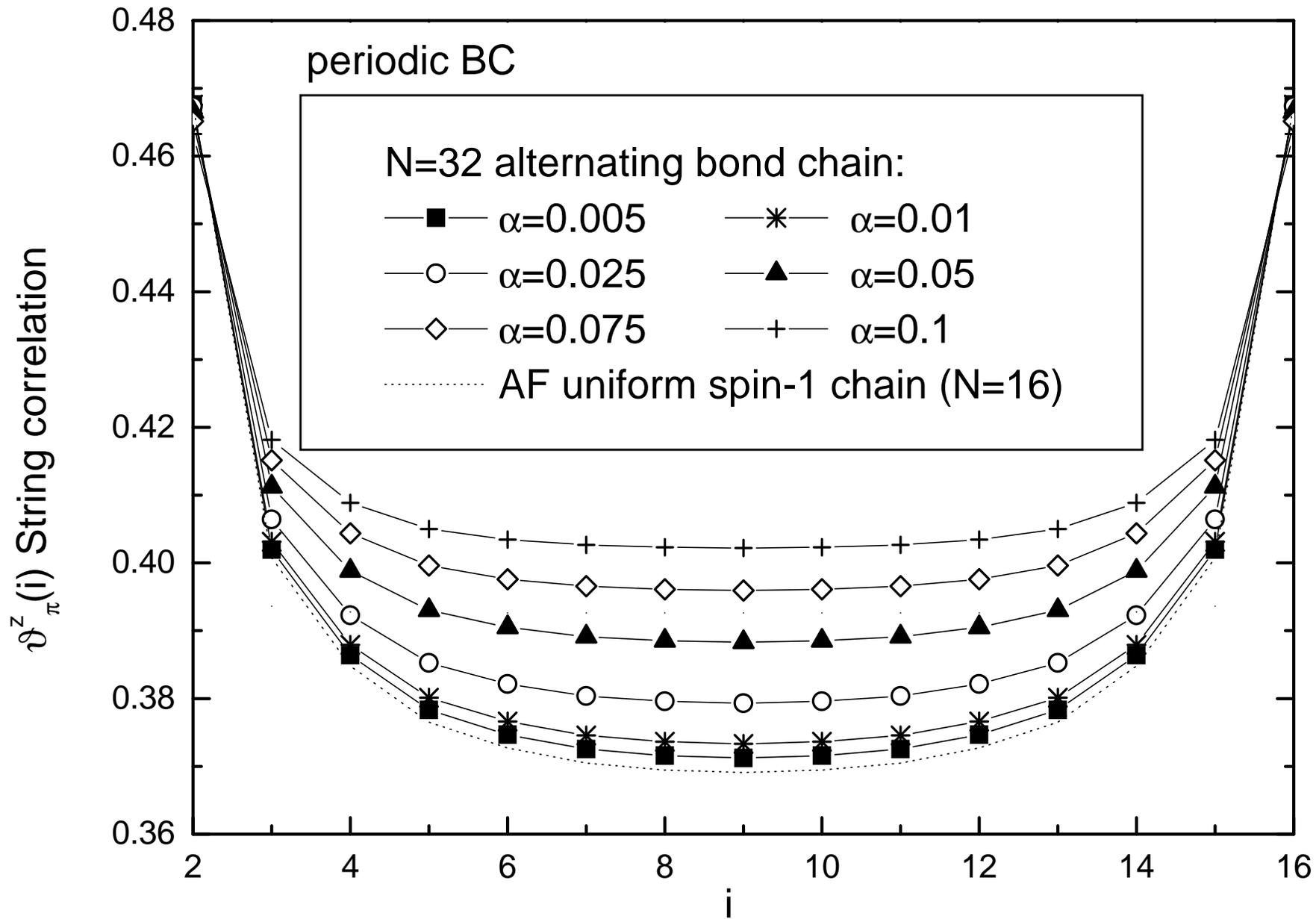
Figure 3:\ The string correlation of the alternating bond chain (N=32) with
periodic boundary condition.

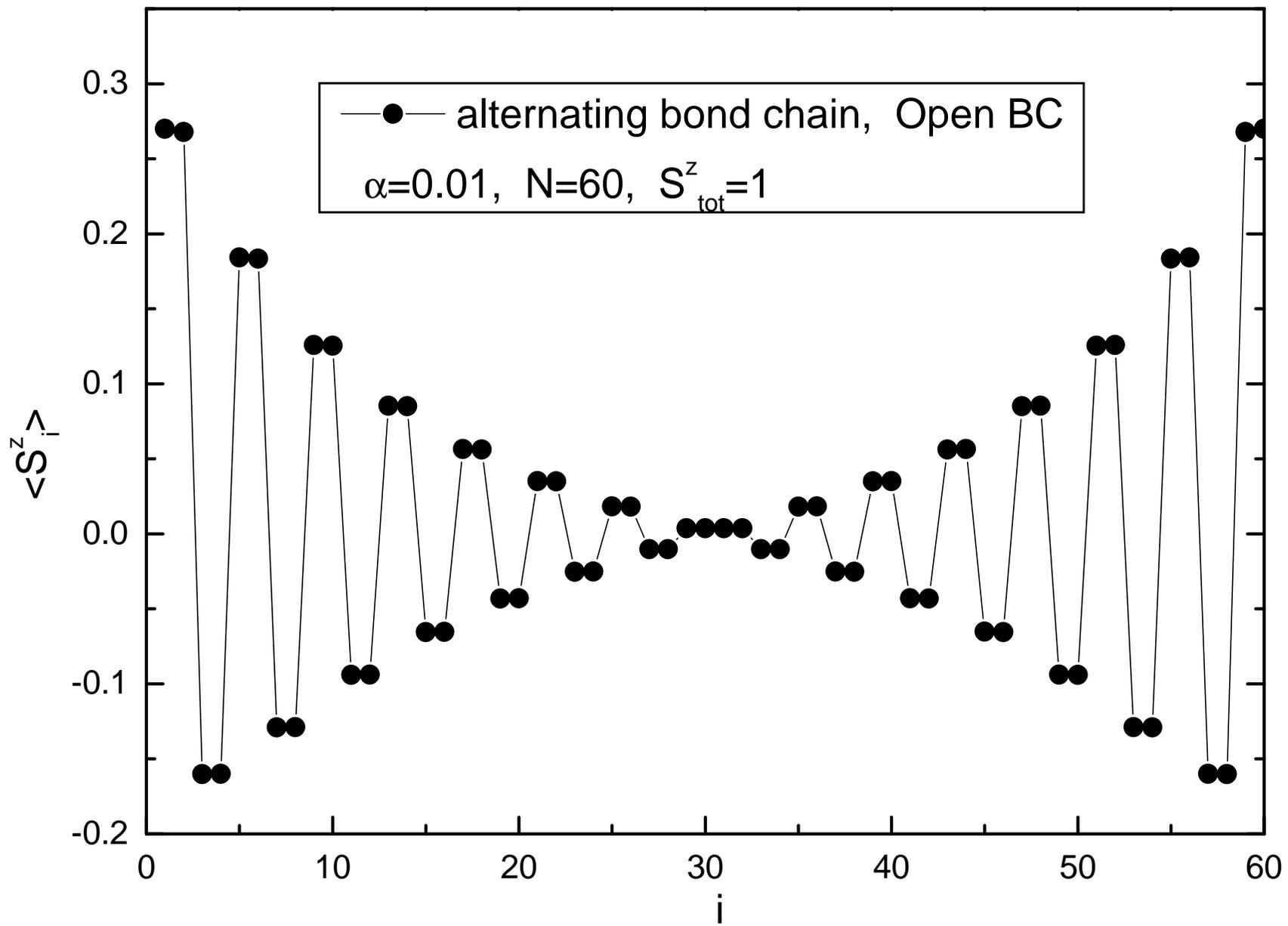
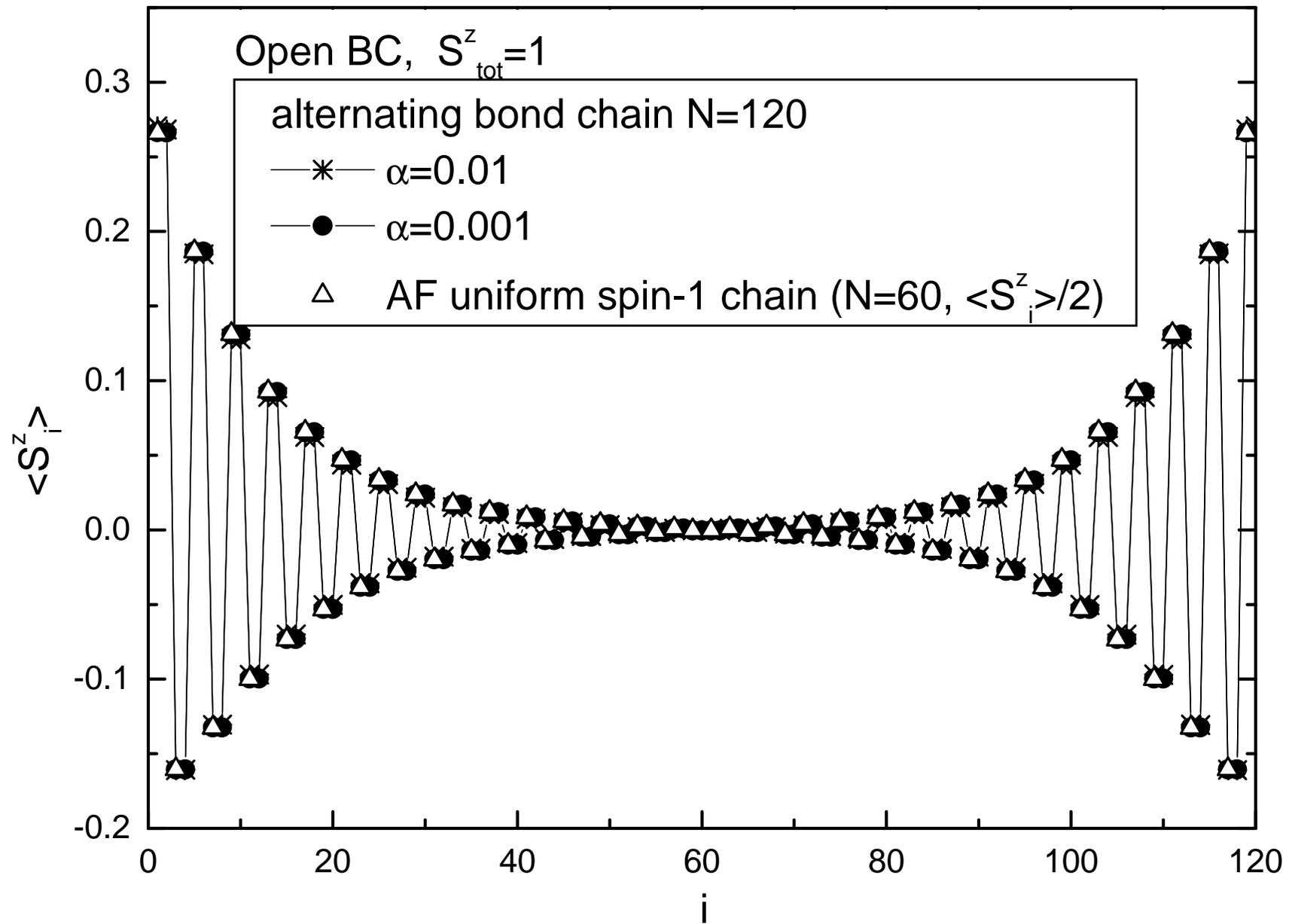
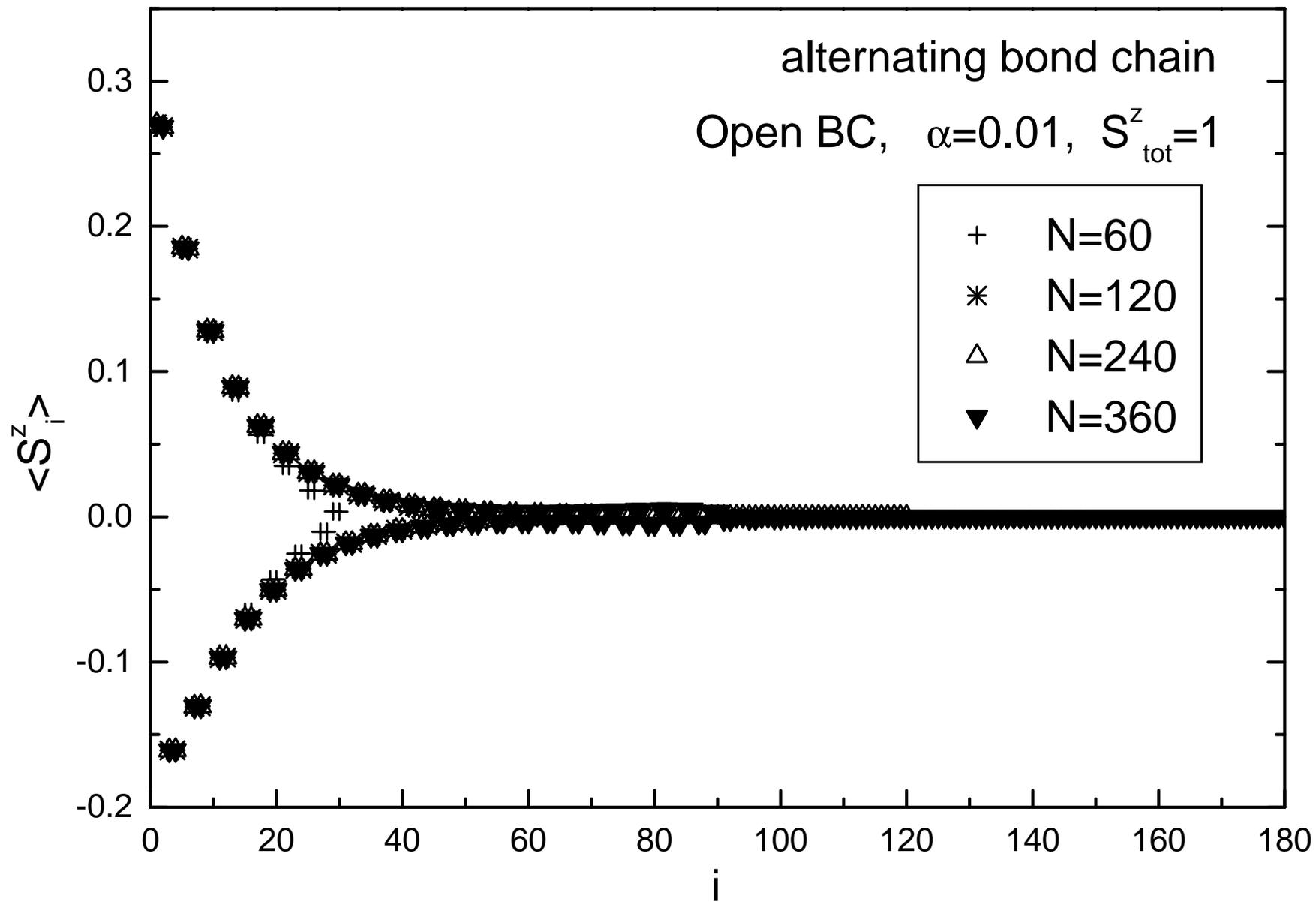
Figure 4: The spin density of an alternating bond chain in subspace
$S_{tot}^{z}=1$ in two cases: (a) $\alpha=0.01$, $N=60$, (b) $\alpha=0.01$ and
$\alpha=0.001$, $N=120$. The data $\langle S_{i}^{z}\rangle/2$ corresponds to
the AFH\ spin-1 chain with N=60. (c) The comparison between several cases with
different numbers of sizes, each locus corresponds to a half of different N.

Figure 5: $\Delta_{1/2}$ is the gap of model (1) with open (or periodic)
boundary condition and $\alpha=0.002$; $\Delta_{1}$ is the Haldane gap of the
AF uniform spin-1 chain with $J=J_{2}=0.002$ (or $0.0005$).

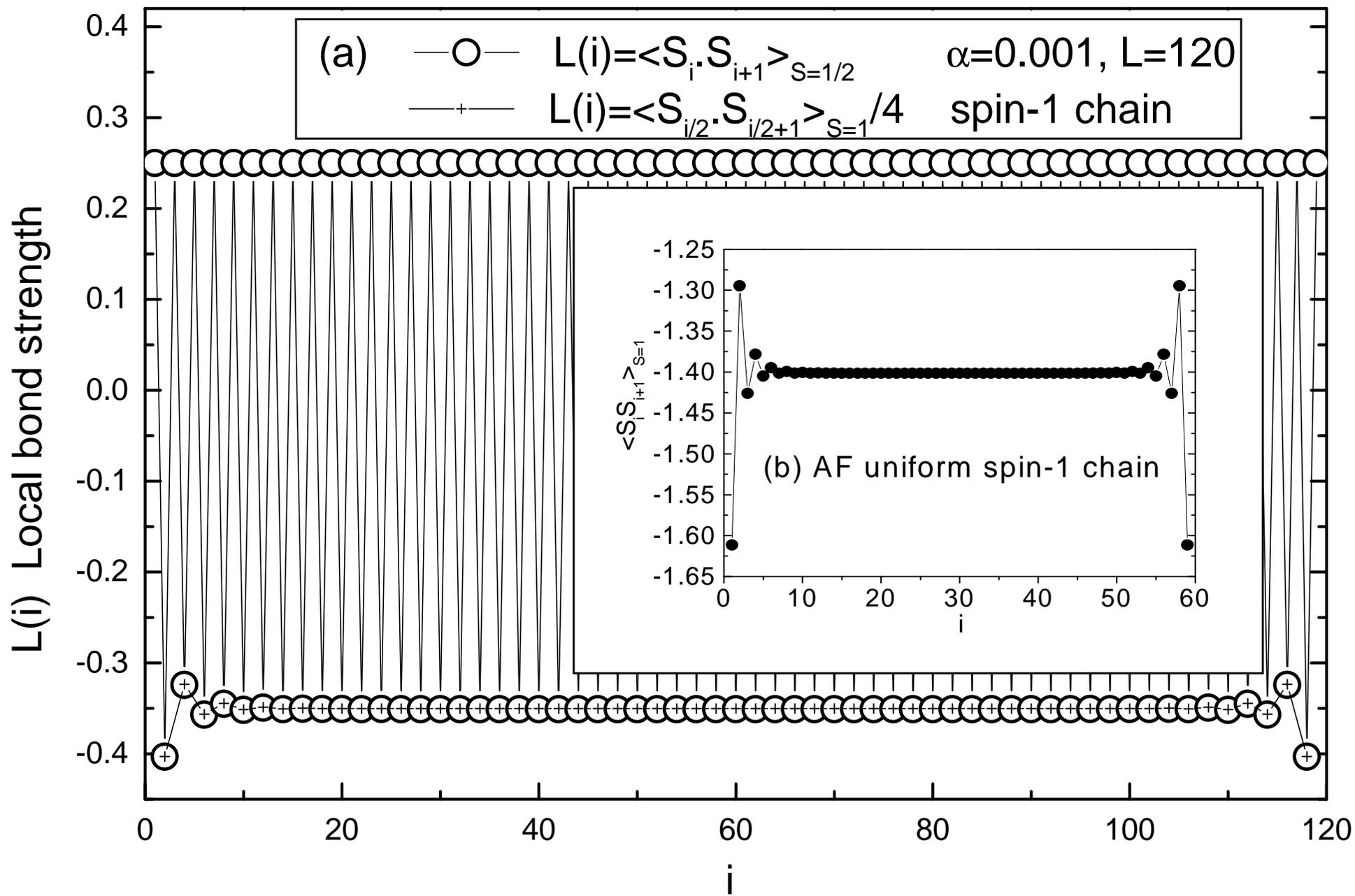
Figure 6: (a) The comparison of the local bond strength $L(i)$ of the ground
state between model (1) of 120 sites with $\alpha=0.001$ and the AF uniform
spin-1 chain of 60 sites. $L(i)$ are $\langle S_{i}\cdot S_{i+1}%
\rangle_{S=1/2}$ and $\langle S_{i/2}\cdot S_{i/2+1}\rangle_{S=1}/4$ for model
(1) and the AFH spin-1 chain, respectively. (b) The local bond strength
$\langle S_{i}\cdot S_{i+1}\rangle_{S=1}$ of the ground state of the AF
uniform spin-1 chain with N=60.

\end{document}